\documentclass[fleqn,12pt,twoside]{article}
\usepackage{espcrc1}
\usepackage{graphics}
\title{Probing hadronization and freeze-out with\\
 multiple strange hadrons and strange resonances}

\author{Marcus Bleicher\address{SUBATECH, Laboratoire de Physique Subatomique 
et des Technologies Associ\'{e}es,\\
University of Nantes - IN2P3/CNRS - Ecole des Mines de Nantes,\\
4 rue Alfred Kastler, F-44072 Nantes Cedex 03, France}
}

\begin{document}

\maketitle


The major goal of the various heavy ion programs is the search for 
a transient state of deconfined matter, dubbed the quark-gluon-plasma (QGP):
A phase transition to this new state of matter is predicted by lattice
QCD when a sufficiently high energy density ($\epsilon 
\approx 1$~GeV/fm$^3$) is reached. 

Strange particle yields and spectra are key  probes
to study excited nuclear matter and to detect 
the transition of (confined) hadronic matter to
quark-gluon-matter (QGP)\cite{qgpreviews,rafelski}. 
The relative enhancement of strange and  multi-strange 
hadrons, as well as hadron ratios in central    
heavy ion collisions with respect to peripheral or proton   
induced interactions have been suggested as a signature    
for the transient existence of a QGP-phase \cite{rafelski}.

A wealth of systematic information has been gathered to study the 
energy dependence of observables from 
$\sqrt s \approx 2$~AGeV to $\sqrt s =200$~AGeV.
For the first time also information on unstable particle emission 
in AA reactions is available: the $\Phi$ and $\Lambda(1520)$ have 
been observed in heavy ion reactions at SPS energies
\cite{NA49Res}, SPS and RHIC 
experiments \cite{NA49Res,STARkstar}  report measurements of 
the $\overline{K^0}(892)$ signal, and RHIC is already attacking
the $f_0$ and $\rho$ mesons and the $\Sigma(1385)$.

A key problem is the identification of  unambiguous signatures of 
possible QGP creation.
Under the assumption of thermal and chemical equilibrium, 
fits with a statistical (thermal) model have been used to 
extract bulk properties of hot and dense matter, e.g. the 
temperature and chemical potential at which chemical freeze-out occurs
\cite{rafelski2,cleymans,braun-munzinger}.
In addition also dynamical evaporation models and non-equilibrium 
transport calculations have been employed to study the energy and 
centrality dependence of particle and especially strangeness 
production\cite{spieles97a,bass98,Soff:1999et,Vance:1999pr}. 
Overall, the conclusions are ambigous and range from evidence for QGP
creation to canonical enhancement and from string fusion to baryon junctions.

Numerous ideas have been brought forward in the
past, ranging from the search for a softest point in the equation of state, 
photon and lepton radiation, $J/\Psi$ suppression to 
event-by-event fluctuations. 
However, the main difficulty in the interpretation of the available data is that 
the observed final state 
hadrons carry relatively little information about their primordial sources.
Most of the hadrons had been subject to many secondary 
interactions and are strongly influenced
by the decays of high mass resonances.
To shed some light on the strangeness production mechanism and the question
of chemical and kinetic equilibration and decoupling, we study
\begin{itemize}
\item
the production of multiple strange baryons in pp interactions. Here
on can directly probe the microscopic decay of color flux tubes,
allowing to differentiate between different string models and
a statistical description of the hadronization.
\item
The energy and centrality dependence of (strange) hadron resonances
in AA which carry unique information about the stage between chemical
and kinetic freeze-out.
\end{itemize}

\paragraph{Hadronization in proton-proton interactions}

In the string picture, high energy proton-proton collisions create
{}``excitations'' in form of strings, being one dimensional objects
which decay into hadrons according to longitudinal phase space. This
framework is well confirmed in low energy electron-positron annihilation
\cite{Werner:1993uh} where the virtual photon decays into a quark-anti-quark
string which breaks up into various kinds of hadrons.
However, specific string models  differ in their
philosophy and the types of strings that are created. In general two
classes of models - based on color or momentum exchange - can be 
distinguished, the resulting objects two quark-diquark strings 
with valence quarks being their ends, however, is quite similar.
Here we will contrast the present UrQMD prescription for 
string formation (similar to the PYTHIA model \cite{pythiamodel} with 
the approach used in NeXuS 3.0:
In UrQMD\cite{urqmdmodel} the projectile and target protons become
excited objects due to the momentum transfer in the interaction.
The resulting strings, with at most two strings being formed,  are
of the di-quark-quark type.
\begin{figure}[h!]
\centerline{\resizebox*{!}{0.2\textheight}{\includegraphics{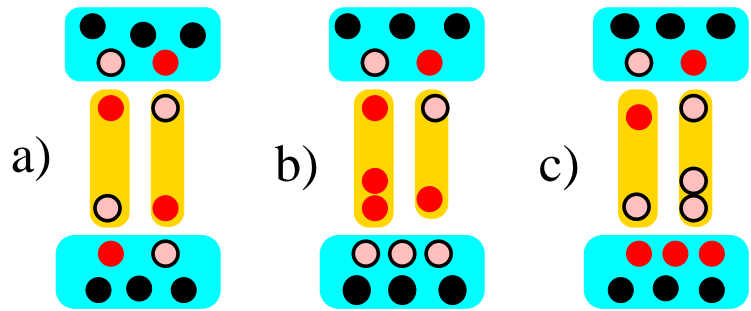}}} 
\vspace*{-.8cm}
\caption{\label{nexus2} {\small a) The simpliest collision configuration
has two remnants and one cut Pomeron represented by 
two \protect\( \mathrm{q}-\overline{\mathrm{q}}\protect \)
strings. b) One of the \protect\( \overline{\mathrm{q}}\protect \)
string-ends can be replaced by a \protect\( \mathrm{qq}\protect \)
string-end. c) With the same probability, one of the \protect\( \mathrm{q}\protect \)
string-ends can be replaced 
by a \protect\( \overline{\mathrm{q}}\overline{\mathrm{q}}\protect \)
string-end (taken from \cite{Liu:2002gw}).}} 
\end{figure}

In NeXuS 3.0\cite{nexusmodel}, the pp interaction is described in
terms of Pomeron exchanges or ladder diagrams. Both hard and soft
interactions take place in parallel. Energy is shared equally
between all cut Pomerons and the remnants. 
Here, for the string ends of soft and semi-hard Pomerons quarks and
anti-quarks from the sea are taken in a flavour-symmetric way. Thus, 
the valence quarks stay in the remnants, yielding excited quark 
bags \cite{Liu:2002gw} (see Fig. \ref{nexus2}).
In contrast to these string models, the predictions of two
statistical models (model {\bf I}, being fully canonical \cite{becalast} and 
model {\bf II}, being canonical with respect to strangeness \cite{ahmed}) 
are also shown.

\begin{figure}[h!]
\vspace{-.2cm}
\centerline{\resizebox*{!}{0.35\textheight}{\includegraphics{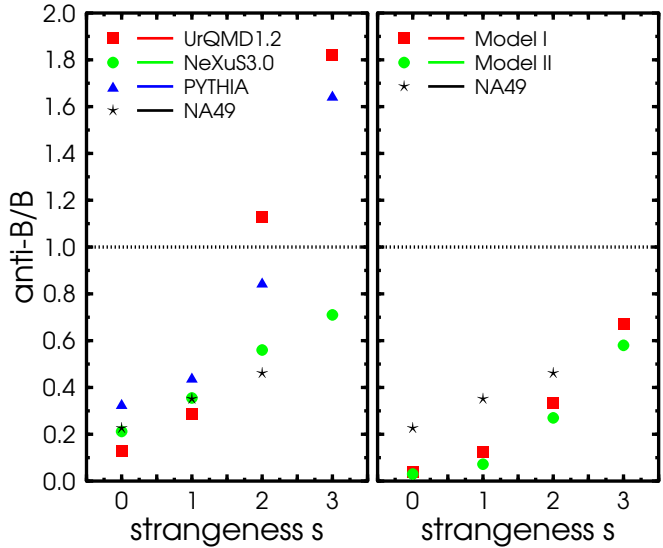}}} 
\vspace*{-1.6cm}
\caption{Left: anti-baryon to baryon ratio at $|y-y_{\rm cm}|<1$ in pp
interactions at 160 GeV as given by PYTHIA, NeXuS3.0 and UrQMD.
Right: anti baryon to baryon ratio for the same reaction as given
by statistical models. Stars depict preliminary NA49 data for the
$\overline B /B$ ratio at midrapidity. \label{pythia}}
\end{figure}

Fig.~\ref{pythia} (left) depicts the anti-baryon to baryon ratio at
midrapidity in proton proton interactions at 160 GeV. The results
of the   calculations by the new NeXuS~3.0 (ratios calculated 
from \cite{Liu:2002gw}) and UrQMD/PYTHIA, which are well established 
string-fragmentation models for elementary hadron
hadron interactions, are included in this figure. In both di-quark
models, the $\overline B /B$ ratio increases strongly with the
strangeness content of the baryon. For strangeness $|s|=3$ the
ratio significantly exceeds unity. In UrQMD and PYTHIA the
hadronization of the di-quark-quark strings leads directly to the
overpopulation of $\overline \Omega$ as discussed in detail 
in \cite{Bleicher:2001nz}. 
The new string formation scheme employed in NeXuS~3.0, however,
allows to get a reasonable agreement with experimental data. 
The basic features of the production of multi-strange
baryons as well as of \( \Lambda  \)s and protons can be understood 
within the model picture of proton-proton scattering: the created final
particles emerge from a non-trivial system of projectile/target remnant 
states and a number of cut Pomerons each represented by a pair of strings.

The predictions of the statistical models (in full phase space) 
are shown in Fig.~\ref{pythia} (right). 
In these approaches the $\overline B /B$ ratio is seen 
to exhibit a significantly weaker  increase with
the strangeness content of the baryon than expected in the di-quark string
fragmentation models. In the grand canonical picture, where 
the $\overline{B}/B$ ratio is very sensitive to the baryon 
chemical potential $\mu_B$, it is easily understood 
that statistical models, can not yield a
ratio of $\overline{\Omega}/\Omega > 1$.  For
finite baryon densities as in the pp system, the $\overline{B}/B$ 
has to stay below one and only in the limit of $\mu_B = 0$ may
$\overline{\Omega}/\Omega = 1$ be approached. This feature is not
blurred in the canonical approach.
For comparison, both figures include
preliminary data on the $\overline B /B$ ratios obtained at
midrapidity by the NA49 Collaboration \cite{na49pp} (Very preliminary NA49 data
seems to support a $\overline\Omega/\Omega$ ratio below one \cite{addendum10}).
\begin{figure}[t]
\vspace{-.2cm}
\centerline{\resizebox*{!}{0.43\textheight}{\includegraphics{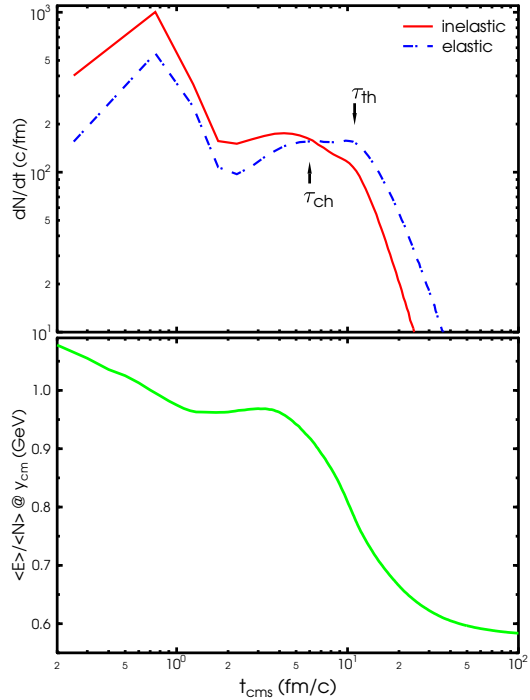}}} 
\vspace*{-1.3cm}
\caption{Top: Inelastic and (pseudo-)elastic collision rates in Pb+Pb at 160AGeV.
$\tau_{\rm ch}$ and $\tau_{\rm th}$ denote the chemical and 
thermal/kinetic freeze-out as given by the microscopic reaction 
dynamics of UrQMD. Bottom: Average energy per particle
at midrapidity ($|y-y_{\rm cm}|\le 0.1$) as a function of 
time (taken from \cite{Bleicher:2002dm}).
\label{chf}}
\end{figure}

It is important to note that the result $\overline{\Omega}/\Omega > 1$
in pp collisions from the di-quark string models solely depends on the geometry 
of the decaying objects. The ratio can not be altered (or even inverted)
by modifications of the strangeness and di-quark suppression factors.
These factors can only modify the absolute yields of strange hadrons, but
do not influence the ratio discussed here.

If the new NA49 data can be confirmed, one is forced to conclude that 
the di-quark string models
fail to describe the ratios of multiple strange baryons in pp
interactions at the SPS. A solution might be to replace the conventional
approach by a system of sea quark strings and remnants or to
abandon the flux tube picture, which has explained many dynamical features of hh
collisions, for a statistical hadronization model.

\noindent
\paragraph{Chemical and kinetic freeze-out in AA collisions}

In nucleus-nucleus collisions the situation is by far more involved 
than in the pp case discussed above. Here, the major fraction of 
finally observed particles in the experimental setup stem from decays of 
resonances (mesonic or baryonic) which have undergone many scatterings
from their point of production to observation. The final hadron yields
seem to be compatible with a hadronic gas described by the 
baryo-chemical potential $\mu_B$ and a temperature parameter $T$ in a 
statistical model (see e.g. \cite{rafelski2,cleymans,braun-munzinger}). 
It has been suggested that both parameters  are coupled by
a universal freeze-out criterion assuming a mean energy per 
hadron $\langle E\rangle/\langle N\rangle = 1$~GeV/hadron \cite{Cleymans:1998fq}
at chemical freeze-out.
In this framework, the formed hadrons do only undergo elastic collision
from this chemical freeze-out time (at SPS energies at a 
temperature of $\approx 160 - 170$~MeV to the final kinetic freeze-out
at temperatures of the order of 120~MeV.

To explore whether this sequential freeze-out is realized 
in heavy ion reactions at highest energies and how it can be probed, 
the Ultra-relativistic Quantum Molecular Dynamics model (UrQMD 1.2)
is used \cite{urqmdmodel}.
This microscopic transport approach is based on the
covariant propagation of constituent quarks and di-quarks accompanied 
by mesonic and baryonic degrees of freedom. As discussed above, 
the leading hadrons of the fragmenting strings contain the valence-quarks 
of the original excited hadron and represent a simplified picture 
of the leading (di)quarks of the fragmenting string. 
The elementary hadronic interactions are modelled according to 
measured cross sections and angular distributions. If the cross 
sections are not experimentally known, detailed balance is employed 
in the energy range of resonances. The partial and total decay widths 
are taken from the Particle Data Group. 

To analyse the different stages of a heavy ion collision, Fig. \ref{chf} (top)
depicts the time evolution of the elastic and inelastic collision rates in central 
Pb+Pb interactions at 160AGeV. The inelastic collision rate (full line) is 
defined as the number of
collisions with flavour changing processes (e.g. $\pi\pi\rightarrow K\overline K$).
The elastic collision rate (dashed dotted line) consists of two components,
true elastic processes (e.g. ${\rm K}\pi\rightarrow {\rm K}\pi$) and pseudo-elastic
processes (e.g. ${\rm K}\pi\rightarrow {\rm K}^* \rightarrow {\rm K}\pi$).
While elastic collision do not change flavour, pseudo-elastic collisions are different.
Here, the ingoing hadrons are destroyed and a resonance is formed. If this resonance
decays later into the same flavours as its parent hadrons, this scattering
is pseudo-elastic.
Fig. \ref{chf} (bottom) depicts the average thermal energy - calculated 
from interacted hadrons with $p^2_z=\left(p^2_x+p^2_y\right)/2$ - 
per particle at midrapidity ($|y-y_{\rm cm}|\le 0.1$).

The present microscopic study of the collision dynamics, indeed supports the
idea of separated phases in the evolution of the system:
\begin{itemize}
\item[{\bf I}] $t<2$~fm/c: In the initial stage of the nucleus-nucleus 
reaction non-equilibrium dynamics leads to strong baryon stopping in
multiple inelastic interactions, shown by huge and strongly time dependent
collision rates. This stage deposits a large amount of (non-thermalized) energy 
and creates the first generation particles.

\item[{\bf II}] 2~fm/c$<t<6$~fm/c: Due to the high particle densities and energies
inelastic scatterings dominate this stage of the reaction.
Chemical equilibrium might be achieved due to a large number of  flavour 
and hadro-chemistry changing processes  until chemical freeze-out.

\item[{\bf III}] 6~fm/c$<t<11$~fm/c: After the system has expanded and cooled down,
elastic and pseudo-elastic collisions take over. Here, only the momenta
of the hadrons change, but the chemistry of the system is mainly unaltered,
leading to the kinetic freeze-out of the system.

\item[{\bf IV}] $t>11$~fm/c: Finally, the reactions cease and the scattering rates
drop drastically. The systems breaks up.

\end{itemize}

Even the hypothesis of chemical decoupling at an energy per hadron of 1~GeV, 
is in line with our analysis as demonstrated in Fig. \ref{chf} (bottom).
In the non-equilibrium stage $\langle E\rangle/\langle N\rangle$ decreases steadily.
While the system is 'cooking' in the inelastic scattering stage, 
$\langle E\rangle/\langle N\rangle$ stays constant around 1~GeV and drops suddenly
as the hadronic system enters the kinetic stage.

The spectra and abundances of $\Lambda(1520)$, $K^0(892)$ and other resonances
can be used to study the break-up dynamics of the source between
chemical and thermal freeze-out.
If chemical and thermal freeze-out are not separated - e.g. due to 
an explosive break-up of the source - all initially produced resonances 
are reconstructable by invariant mass analysis of the final state hadrons. 
However, if there is a separation between the different freeze-outs, 
a part of the resonance daughters rescatter, making this
resonance unobservable in the final state. Thus, the relative suppression of
resonances in the final state compared to those expected from thermal
estimates provides a chronometer for the time period between the different
reaction stages. 
Even the chemical composition of the system might be changed by up to 10\%
in the hyperon sector (after chemical 'freeze-out'), due to
inelastic  scatterings of the resonance daughters 
(e.g. $\overline K p \rightarrow \Lambda$). 

\begin{figure}[h!]
\vspace{-.2cm}
\centerline{\resizebox*{!}{0.35\textheight}{\includegraphics{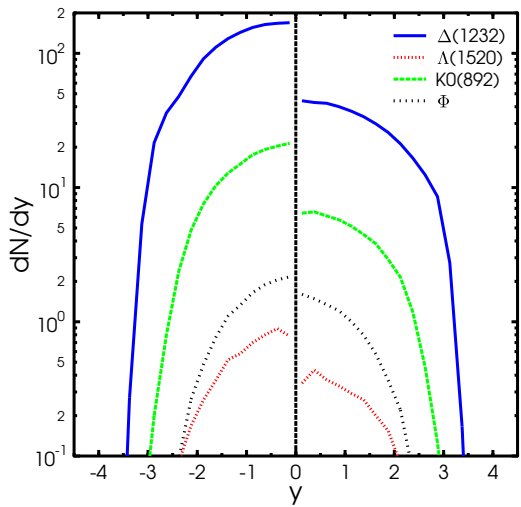}}} 
\vspace*{-1.3cm}
\caption{Rapidity densities for $\Delta(1232)$, $\Lambda^*(1520)$, K$^0(892)$
and $\Phi$ in Pb($160\,A$GeV)Pb, b$<3.4$~fm collisions.
Left: All resonances as they decay. Right: Reconstructable 
resonances (taken from \cite{Bleicher:2002dm}).
\label{dndy}}
\end{figure}

The rapidity spectra for $\Delta(1232)$, $\Lambda^*(1520)$, K$^0(892)$
and $\Phi$ in Pb($160\,A$GeV)Pb, b$<3.4$~fm collisions.
are depicted in Fig. \ref{dndy}.
The left part, shows the total amount of decaying resonances.
Here, subsequent collisions of the decay products have
not been taken into account - i.e. whenever a resonance decays during
the systems evolution it is counted.
However, the additional interaction of the daughter hadrons
disturbs the signal of the resonance in the invariant mass spectra.
This lowers the observable yield of resonances drastically as
compared to the primordial yields at chemical freeze-out.
Fig. \ref{dndy} (right) addresses this in the rapidity distribution of those
resonances whose decay products do not suffer subsequent collisions -
these resonances are in principle reconstructable from their decay products.
Note that reconstructable in this context still assumes reconstruction of
all decay channels, including many body decays.

By using the estimates done by \cite{Torrieri:2001ue} in a statistical
model, it seems possible to relate the result of the present microscopic
transport calculation to thermal freeze-out parameters.
The surprising result is that the microscopic source seems to have 
a lifetime shorter than 1~fm/c and a freeze-out
temperature lower than 100~MeV. Apparently, the values obtained 
from UrQMD seem to favour a scenario of a sudden
break-up of the initial hadron source, in contrast to the time evolution
of the chemical and thermal decoupling as shown in Fig. \ref{chf}.
This misleading interpretation might be due to the re-creation
of resonances in the elastic scattering stage, which was not taken
into account in the statistical model analysis. The influence of
these effects is currently under investigation.

However, even the question of the existence of
such resonance states in the hot and dense environment is still
not unambiguously answered. Since hyperon resonances are  
expected to dissolve at high energy densities (see .e.g. \cite{Lutz:2001dq})
it is of utmost importance to study the cross section
of hyperon resonances as a thermometer of the collision.

\begin{figure}[h!]
\centerline{\resizebox*{!}{0.3\textheight}
{\includegraphics{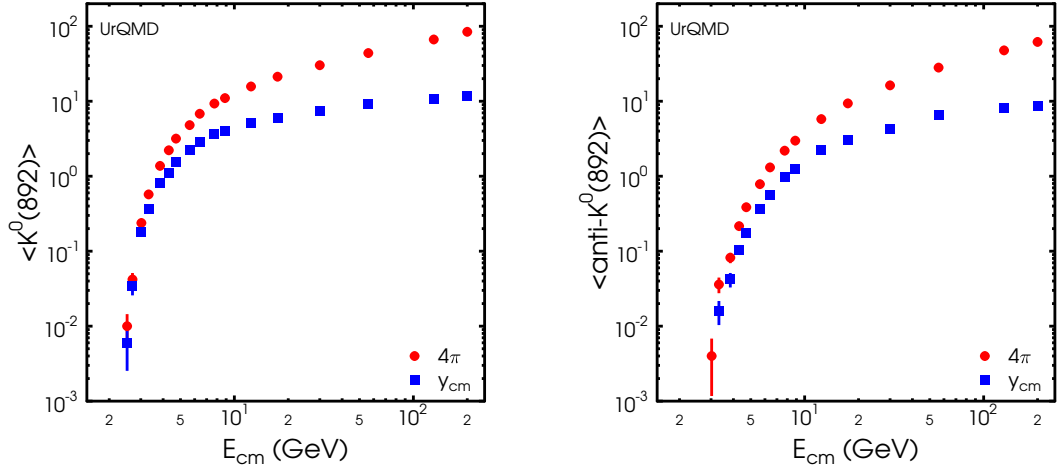}}}
\caption{Multiplicity excitation function for central Pb+Pb (Au+Au) reactions, 
in $4\pi$ (circles) and at midrapidity (squares). Left: $\langle K^0(892)\rangle$.
Right: $\langle \overline{K^0}(892)\rangle$. 
\label{exc.kstar}}
\end{figure}

\begin{figure}
\centerline{\resizebox*{!}{0.3\textheight}
{\includegraphics{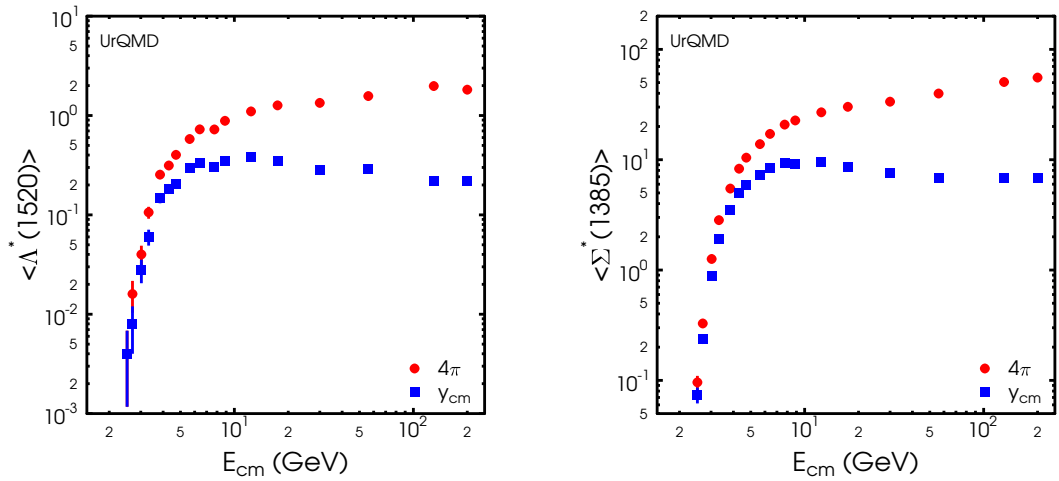}}}
\caption{Multiplicity excitation function for central Pb+Pb (Au+Au) reactions, 
in $4\pi$ (circles) and at midrapidity (squares). 
Left: $\langle \Lambda^*(1520)\rangle$.
Right: $\langle \Sigma^*(1385)\rangle$ (all charges summed). 
\label{exc.lstar}}
\end{figure}

To set the stage, Figs. \ref{exc.kstar} and \ref{exc.lstar} address 
the excitation function of observable resonance multiplicities. 
In addition, a comparison of rapidity 
integrated yields ($4\pi$ values, circles) with the center-of-mass 
values ($y_{\rm cm}\pm 0.5$, squares) is given. 
The $\Lambda$ includes decays from $\Sigma^0$, but not from $\Xi$. 
Protons do not include decays from $\Lambda$'s.
All strong decays are included. No cuts are applied except when mentioned.
The anti-K$^*$ multiplicities are monotonously increasing with energy, 
while the hyperon resonance show a pronounced maximum in the excitation function 
at $\sqrt s=$8~GeV (E$_{\rm lab} \approx 30$~AGeV). 
This maximum is also present - as earlier observed for  
hyperons by \cite{Braun-Munzinger:2001as} - if scaled by the number of pions.
This makes the newly planned GSI200 facility an ideal place to study
in-medium modifications of resonances.

\begin{figure}
\centerline{\resizebox*{!}{0.3\textheight}
{\includegraphics{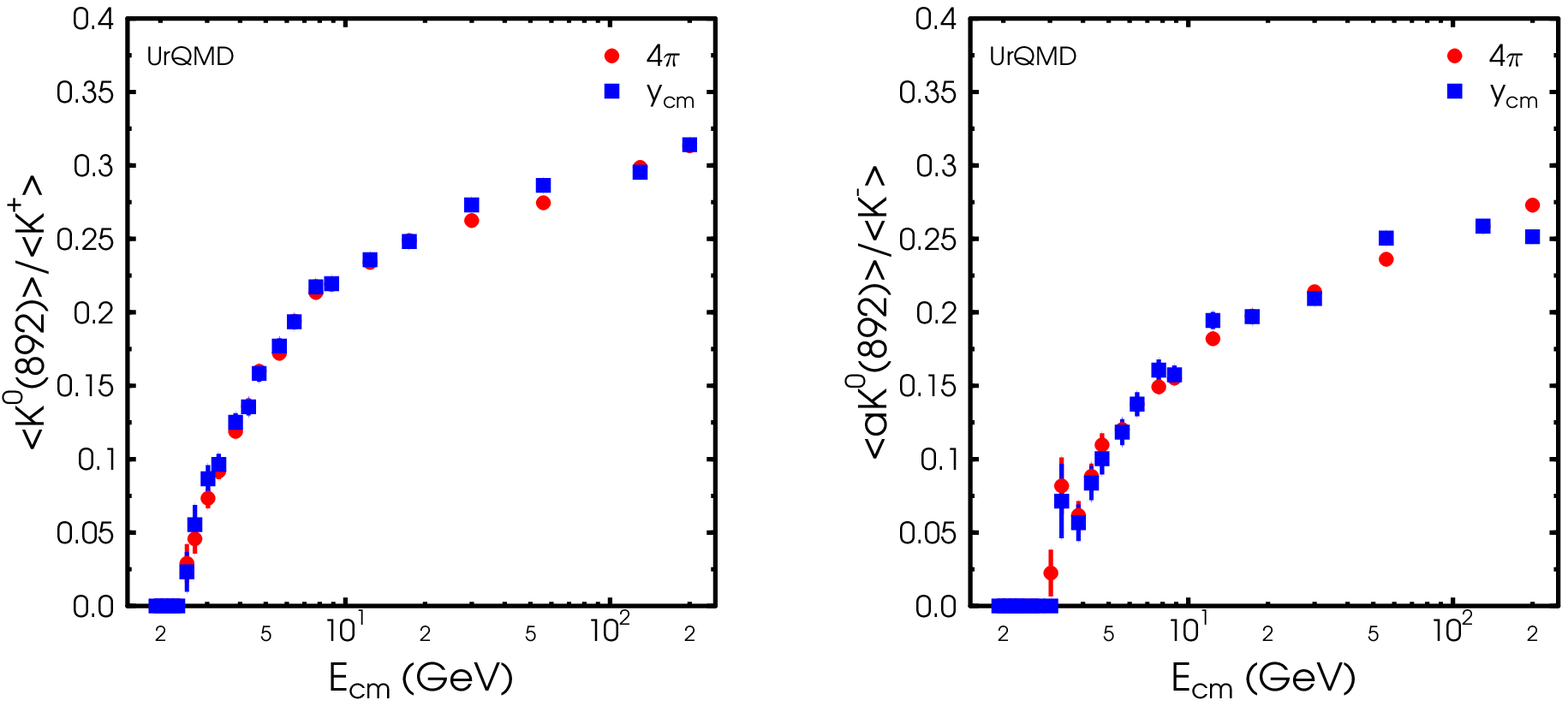}}} 
\caption{Ratio excitation function for central Pb+Pb (Au+Au) reactions, 
in $4\pi$ (circles) and at midrapidity (squares). 
Left: $\langle K^0(892)\rangle /\langle K^+\rangle$.
Right: $\langle \overline{K^0}(892)\rangle /\langle K^-\rangle$. 
\label{exc.kstaroverk}}
\end{figure}

\begin{figure}
\centerline{\resizebox*{!}{0.3\textheight}
{\includegraphics{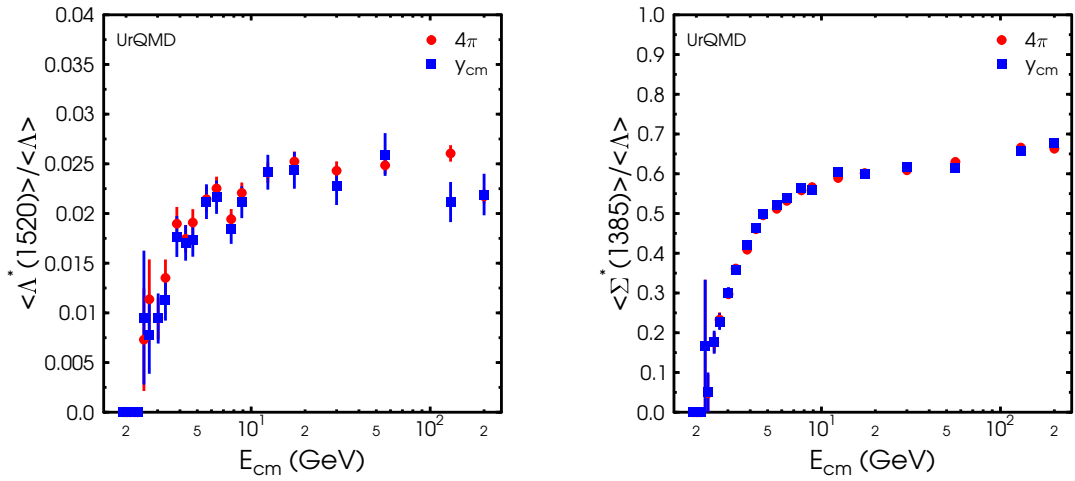}}} 
\caption{Ratio excitation function for central Pb+Pb (Au+Au) reactions, 
in $4\pi$ (circles) and at midrapidity (squares). 
Left: $\langle \Lambda^*(1520)\rangle /\langle \Lambda\rangle$.
Right: $\langle \Sigma^*(1385)\rangle /\langle \Lambda\rangle$. 
\label{exc.lstaroveral}}
\end{figure}

More information can be obtained if the multiplicities are normalised
to the groundstate hadrons, i.e. (anti-)Kaons and Lambdas.
Figs. \ref{exc.kstaroverk} and \ref{exc.lstaroveral} show the
energy dependence of the $h^*/h$ ratios. Here the baryo-chemical potentials
cancel out and information on the freeze-out temperature can be obtained.

To summarise, hadronization into multiple strange hadrons in 
elementary pp interactions  has been studied within different string 
approaches and statistical models and confronted with data. 
Di-quark string models seem not to be in line with the experimental observations
of $\overline\Omega$ and $\Omega$ production in pp at the SPS.
The freeze-out dynamics in AA has been microscopically explored. 
Different decoupling stages have been identified in the 
non-equilibrium dynamics. Observables to clock the kinetic scattering stage
and the freeze-out temperature with (strange-) resonances were discussed.

\section*{Acknowledgements}
I want to acknowledge interesting and stimulating discussions with
Drs. Klaus Werner, J\"org Aichelin, Fuming Liu, Tanguy Pierog, Christina Markert 
and Giorgio Torrieri. 

\clearpage




\begin{thebibliography}{99}

\bibitem{qgpreviews}
S.~A.~Bass, M.~Gyulassy, H.~St\"ocker and W.~Greiner,
J.\ Phys.\ {\bf G25}, R1 (1999); 
Stock~R, Phys. Lett. {\bf B456}, (1999) 277

\bibitem{rafelski} J.~Rafelski and B.~M\"uller,
Phys. Rev. Lett. {\bf 48}, 1066 (1982); J.~Rafelski, Phys. Rep.
{\bf 88}, 331 (1982); P.~Koch, B.~M\"uller and J.~Rafelski,
 Phys. Rep. {\bf 142}, 167 (1986).


\bibitem{NA49Res}
Ch. Markert, PhD thesis, Univ. Frankfurt; 
V. Friese [NA49 collaboration],  http://www.rhic.bnl.gov/qm2001/

\bibitem{STARkstar}
Z.~b.~Xu  [STAR Collaboration],
Nucl.\ Phys.\ A {\bf 698} (2002) 607
[arXiv:nucl-ex/0104001].
For the latest results, see these proceedings.


\bibitem{rafelski2}
J.~Letessier,  A.~Tounsi, U.~Heinz, J.~Sollfrank and J.~Rafelski,
Phys. Rev. Lett. {\bf 70}, 3530 (1993); J.~Letessier, J.~Rafelski
and A.~Tounsi, Phys. Lett. {\bf B321}, 394 (1994); J.~Rafelski and
M.~Danos, Phys. Rev. {\bf C50}, 1684 (1994).

\bibitem{cleymans}
J. Cleymans and K. Redlich,
 Phys. Rev. Lett. {\bf 81}, 5284 (1998);
F.~Becattini, J.~Cleymans, A.~Keranen, E.~Suhonen and K.~Redlich,
Phys.\ Rev.\  {\bf C64}, 024901 (2001).


\bibitem{braun-munzinger}
P.~Braun-Munzinger, J.~Stachel, J.~P. Wessels and N.~Xu, Phys.
Lett. {\bf B344}, 43 (1995); 
{\bf B365}, 1 (1996); 
P. Braun-Munzinger, I. Heppe  and J. Stachel, Phys.
Lett.  {\bf B465}, 15 (1999);
P.~Braun-Munzinger,
 D.~Magestro, K.~Redlich and J.~Stachel,
Phys.\ Lett.\  {\bf B518}, 41 (2001).

\bibitem{spieles97a}
C.~Spieles, H.~Stocker and C.~Greiner,
Eur.\ Phys.\ J.\  {\bf C2} 351 (1998) 351;
C.~Greiner and H.~St\"ocker., Phys. Rev. {\bf D44}, 3517 (1992).

\bibitem{bass98}
S.~A.~Bass { et al.},
Phys.\ Rev.\ Lett.\  {\bf 81}, 4092 (1998).

\bibitem{Soff:1999et}
S.~Soff, S.~A.~Bass, M.~Bleicher, L.~Bravina, E.~Zabrodin, H.~Stocker and W.~Greiner,
Phys.\ Lett.\ B {\bf 471} (1999) 89
[arXiv:nucl-th/9907026].

\bibitem{Vance:1999pr}
S.~E.~Vance and M.~Gyulassy,
Phys.\ Rev.\ Lett.\  {\bf 83} (1999) 1735

\bibitem{Werner:1993uh} 
K.~Werner, Phys.\ Rept.\ {\bf 232} (1993) 87.

\bibitem{pythiamodel}
H.-U. Bengtsson and T. Sj\"ostrand, { Comput. Phys. Commun.} {\bf
46} 43 (1987).

\bibitem{urqmdmodel}
M.~Bleicher { et al.},
J.\ Phys.\  {\bf G25} 1859 (1999);
S.~A.~Bass { et al.},
Prog.\ Part.\ Nucl.\ Phys.\  {\bf 41} 225 (1998).

\bibitem{nexusmodel}
H.J. Drescher { et al.}, Phys. Rep. {\bf 350} 93 (2001).

\bibitem{becalast}
F. Becattini and  G. Passaleva, hep-ph/0110312, Eur. Phys. J. in print.

\bibitem{ahmed}J. S. Hamieh, K. Redlich and A. Tounsi,
Phys. Lett. {\bf B486} 61 (2000); J. Phys. {\bf G27}, 413 (2001).


\bibitem{Liu:2002gw}
F.~M.~Liu, J.~Aichelin, M.~Bleicher, H.~J.~Drescher, S.~Ostapchenko, T.~Pierog and K.~Werner,
arXiv:hep-ph/0202008.

\bibitem{Bleicher:2001nz}
M.~Bleicher {\it et al.},
Phys.\ Rev.\ Lett.\  {\bf 88} (2002) 202501
[arXiv:hep-ph/0111187].

\bibitem{na49pp} 
T. Susa et al, NA49, Nucl. Phys. A698 (2002) 491c; 
K.~Kadija, talk given at {\it Strange Quark Matter 2001,} Frankfurt, Germany.

\bibitem{addendum10}
NA49 Collaboration, Proposal CERN/SPSC/P264, Addendum 10


\bibitem{Cleymans:1998fq}
J.~Cleymans and K.~Redlich,
Phys.\ Rev.\ Lett.\  {\bf 81} (1998) 5284
[arXiv:nucl-th/9808030].

\bibitem{Bleicher:2002dm}
M.~Bleicher and J.~Aichelin,
Phys.\ Lett.\ B {\bf 530} (2002) 81
[arXiv:hep-ph/0201123].

\bibitem{Torrieri:2001ue}
G.~Torrieri and J.~Rafelski,
Phys.\ Lett.\ B {\bf 509} (2001) 239
[arXiv:hep-ph/0103149].

\bibitem{Lutz:2001dq}
M.~F.~Lutz and C.~L.~Korpa,
arXiv:nucl-th/0105067.

\bibitem{Braun-Munzinger:2001as}
P.~Braun-Munzinger, J.~Cleymans, H.~Oeschler and K.~Redlich,
Nucl.\ Phys.\ A {\bf 697} (2002) 902
[arXiv:hep-ph/0106066].


\end{thebibliography}
\end{document}